Review Article

# Current Landscape of Mesenchymal Stem Cell Therapy in COVID Induced Acute Respiratory Distress Syndrome

**Adrita Chanda¹, Adrija Aich¹, Arka Sanyal², Anantika Chandra³ and Saumyadeep Goswami³***

*¹School of Biosciences and Technology (SBST), Vellore Institute of Technology, Vellore, India*

*²School of Biotechnology, KIIT University, India*

*³School of Bioscience, Indian Institute of Technology, Kharagpur, Kharagpur, West Bengal, India*

***Corresponding Author:** Saumyadeep Goswami, School of Bioscience, Indian Institute of Technology, Kharagpur, Kharagpur, West Bengal, India.

## Abstract

The severe acute respiratory syndrome coronavirus 2 (SARS-CoV-2) outbreak in China's Hubei area in late 2019 has now created a global pandemic that has spread to over 150 countries. In most people, COVID-19 is a respiratory infection that produces fever, cough, and shortness of breath. Patients with severe Covid- 19 may develop ARDS. MSCs can come from a number of places, such as bone marrow, umbilical cord, and adipose tissue. Because of their easy accessibility and low immunogenicity, MSCs were often used in animal and clinical research. In recent studies, MSCs have been shown to decrease inflammation, enhance lung permeability, improve microbial and alveolar fluid clearance, and accelerate lung epithelial and endothelial repair. Furthermore, MSC-based therapy has shown promising outcomes in preclinical studies and phase 1 clinical trials in sepsis and ARDS. In this paper, we posit the therapeutic strategies using MSC and dissect how and why MSC therapy is a potential treatment option for COVID-19-induced ARDS. We cite numerous promising clinical trials, elucidate the potential advantages of MSC therapy for Covid-19 ARDS patients, examine the detriments of this therapeutic strategy and suggest possibilities of subsequent research.

## Introduction

Coronaviruses are non-segmented, positive-sense single-stranded RNA viruses. These enveloped viruses belong to the Coronaviridae family and are known to infect a wide variety of vertebrates [1]. In December 2019, Wuhan, China reported its first case of a disease, causing pneumonia-like symptoms. Initially known as the 2019 Coronavirus novel (1999-nCoV), it was shortly renamed SARS-CoV-2. SARS-CoV-2 shared an approximate 79.5% of gene homology with the SARS-CoV-1 virus genome, which was responsible for the 2002-2004 SARS outbreak [2]. On January 30, 2020, WHO declared the SARS-CoV-2 outbreak as a public health emergency of international concern, and on March 11, 2020, WHO declared the outbreak as a pandemic [3]. The typical symptoms of SARS-CoV-2 often occur between 2-14 days after exposure [4] and it has a mean incubation period of 5.1 days. On average, it has been noted that most people infected with the SARS-CoV-2 virus can recover completely within a few weeks of infection. However, the victims of long- COVID continue to experience discomfort and symptoms even post-recovery from the disease. Most of the damage that COVID-19 inflicts is on the alveoli of the lungs. The scar tissues that result from that damage have been reported to cause long-term breathing problems in COVID-19 patients [5].

At present, most therapeutic strategies for SARS-CoV-2 infected patients are focused primarily on palliative care. Antivirals like

Remdesivir and immune modulators like tocilizumab, combat the onset and persistence of the "cytokine storm" which may lead to hyper-inflammation. In cases where the disease progresses to ARDS, muscle relaxants, corticosteroids like methylprednisolone, and ventilators are used to support the patient. Evidence has shown that stem cell therapies, particularly the use of Mesenchymal Stem Cells (MSCs) in the treatment of ARDS have provided satisfactory results and have proven to be an effective treatment [6,7]. They have also shown promise as a therapeutic strategy in clinical trials, inhibiting over- active immune response, preventing cytokine storm, ameliorating lung damage, and restoring lung function in patients suffering from COVID-19 [8]. MSCs have also been used for disease modelling to elucidate the underlying mechanisms of disease progression in COVID-19. Novel strategies to combat COVID 19 and its detrimental consequences are the need of the world right now. In this review, we aim to conglomerate MSC- based strategies that have the potential to become a therapeutic modality for COVID 19 in near future, as well as to highlight important perspectives related to it.

## Pathogenesis of SARS CoV-2 in human lungs

### Attachment and entry of SARS CoV2 in human lung alveolar cells

The SARS Cov-2 virus initially infects the ciliated and secretory cells present in the nose and nasal cavity which cannot produce vigorous innate immune responses, and thus the virus utilizes the nasal cavity as its storehouse. As the disease progresses, the ciliated cells of the bronchioles and the bronchi are affected. The alveolar surface, bronchial cells, and tracheal epithelial cells have abundant Angiotensin-converting enzyme II (ACE2) receptors [9]. Interestingly, ACE2, which is a part of the Renin-Angiotensin-Aldosterone System (RAAS), has been found to be the receptor for the SARS-CoV-2 virus [10]. These receptors are responsible for lowering blood pressure and also regulate pulmonary oedema by inactivating angiotensin II synthesized by ACE [9].

The SARS-Cov-2 virus uses a surface glycoprotein deemed as the spike protein to bind to the host's ACE-2 receptor present on the surface of the lung epithelial cells [10]. The spike proteins are trimers containing three S1 heads atop a trimeric S2 stalk. The S1 protein is known to contain the receptor-binding domain which is responsible for recognizing and binding to the host's ACE-2 receptor (Figure 1a, Figure 1b).

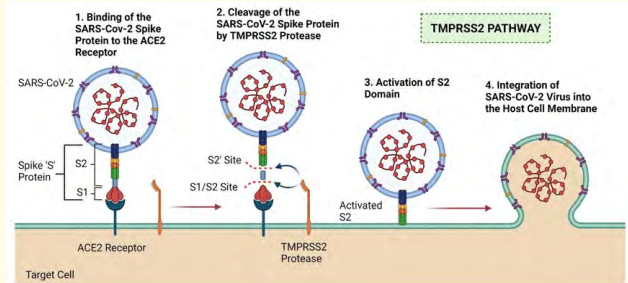

**Figure 1a:** The steps in the TMPRSS2 Pathway of SARS-CoV-2 entrance are mentioned as follows: (1) Initially, the SARS-CoV-2 Spike (S) protein binds to the host ACE2 receptor. (2) The TMPRSS2 protease cleaves the SARS-CoV-2 Spike protein at S1/S2 and S2′ locations, causing the (3) activation of the SARS-CoV-2 Spike protein's S2 Domain, which eventually results in the (4) fusion of SARS-CoV-2 with host cell membranes, which facilitates viral entry. [Adapted from "Mechanism of SARS-CoV-2 Viral Entry", by BioRender.com (2020). Retrieved from https://app.biorender.com/biorender-templates].

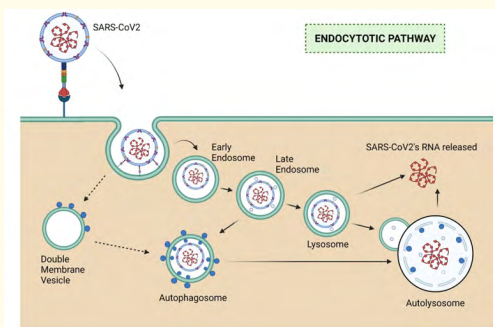

**Figure 1b:** The endocytic route is primarily responsible for SARS-entrance CoV-2's into host cells. In the meantime, autophagy has been linked to viral replication in cells, a process that is partly linked to the development of DMVs (Double Membrane Vesicles) in the host cells. After connecting to the ACE2 receptor, SARS-CoV-2 can be immediately endocytosed into the host cell, forming an Early Endosome, a Late Endosome, and finally a Lysosome. The creation of DMVs, which fuse with the Late Endosome to form the Autophagosome, can result from the virus's endocytosis. After fusion, the Autophagosome and the Lysosome form the Autolysosome. SARS-CoV-2 can manipulare the autophagy process to ensure its survival, resulting in viral replication and the release of SARS-CoV-2 RNA [Created with BioRender.com.].

For the SARS-CoV-2 virus to fuse its membrane to the host's cell, the viral spike protein must be proteolytically cleaved at the junction of its S1 and S2 domains. This cleavage induces a conformational change in the S2 domain that allows the virus to insert its N-terminus into the host cell membrane to enable the fusion of the viral and the host's cellular membranes [10].

The cleavage at the S1/S2 junction is mediated by the availability of a suitable protease. Based on availability, the SARS-CoV-2 virus has been known to utilize two different pathways for entering the host cell. The "early pathway" is dependent on the presence of the cell surface serine-protease TMPRSS2, which enables viral fusion at the cell surface of the host cell [10]. The "late pathway" is dependent on the endocytic uptake of the ACE-2 bound SARS-CoV-2 virus. The presence of cathepsins in the endosomal-lysosomal compartments facilitates the release of the viral genome into the cytoplasm.

**Physiology of a normal human lung:**

In an uninjured lung, the interalveolar septa are the structural platforms for gaseous exchange [11]. The interalveolar septa, surrounded on either side by the alveolar epithelium, refer to the alveolar wall shared by two contiguous alveoli and are made up of interwoven connective tissue fibres. The alveolar epithelium is a monolayer composed of two distinct types of cells – alveolar type I (ATI) and alveolar type II (ATII) cells (Figure 2a). The ATI cells are flat and responsible for mediating oxygen transfer and carbon dioxide excretion. The ATII cells are cuboidal and secrete the surfactant responsible for reducing surface tension and ensuring that the alveoli remain open to facilitate gaseous exchange. Additionally, the ATII cells are also credited with repair and renewal functions during injury and are the resident progenitor cells of the alveoli [11]. In a functional lung, the interstitial fluid is unable to cross into the airspaces of the alveoli because of the tight junctions interlinking the ATI and ATII cells. This causes them to form a continuous barrier selective to the trans vascular flux of solutes, fluid, and inflammatory cells into the alveolar lumen from the interstitial space. The Sodium channels and Sodium-Potassium ATPase pumps present on the alveolar epithelium help reabsorb the excess oedematous fluid which may have leaked into the alveolar lumen from the interstitial spaces. Hence, under normal conditions, the intact alveolar epithelium maintains dry airspaces.

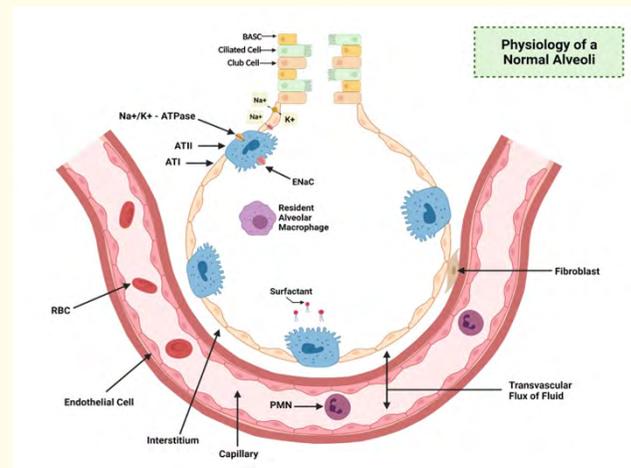

**Figure 2a:** Our alveolar epithelium is a continuous monolayer made up of alveolar type I (ATI) cells, which are very thin cells that allow for gas exchange, and alveolar type II (ATII) cells, which are primarily responsible for producing surfactant molecules that allow for lung expansion with low surface tension. By aiding the transfer of ions and fluid from the alveolus, both cell types help to maintain dry airspaces inside the alveolus. The intercellular tight junction connects the entire alveolar epithelium and is principally responsible for barrier function and regulating fluid and ion transport across the epithelium. Endothelial cells are the cells that make up the alveolar capillary and are in charge of regulating fluid input and the entry of inflammatory cells into the interstitial space. Intercellular junctions, which include both tight and adherens junctions, join these endothelial cells. Small solutes and water are moved into the interstitial space and then into the lymphatic system by the transvascular flux of fluid out of the alveolar capillary in normal circumstances. This fluid cannot cross the epithelial cell barrier in healthy normal alveoli, and resident alveolar macrophages inhabit the alveolar airspaces, providing host defence. Fibroblasts are present in the interstitial space, whereas PMNs and RBCs are situated within the alveolar capillaries. During the commencement of any infection, these PMNs can quickly migrate to the airspaces to carry out an immune response. [Created with BioRender.com.].

## Physiology of a lung in SARS CoV2 infection

One of the primary immune responses employed by the body against viral infections is the employment of reactive oxygen species (ROS). The epithelial cells of the lung, and its resident alveolar macrophages produce ROS by an active endogenous expression of ROS-generating enzymes like xanthine oxidases (Xo) and nicotinamide adenine dinucleotide phosphate oxidases (Nox) (Figure 2b). In healthy lungs, the excess production of ROS on the components of the host's cells is neutralized by the action of endogenously produced antioxidants like superoxide dismutase (SOD) [12]. However, in COVID-19 patients, the neutralization of ROS fails to happen, resulting in the accumulation of ROS in the lung cells of the patient. This anomaly is controlled by the TLR7-MyD88 signaling pathway which results in Nox-2 activation and ROS production, the TLR4-TRIF-TRAF6-NF-kB signaling pathway which is activated by oxidized phospholipids as a consequence of ROS and also, as a result of endoplasmic reticulum (ER) stress caused by an overload of protein production in the cells during SARS-CoV-2 infection, etc. [13]. The virus and its pattern-associated molecular patterns (PAMPs) are sensed by different sensors known as pattern recognition receptors(PRRs). The major PRRs involve TLRs, NLRs, and RLRs. TLR4, TL3, TL7/8, and NLRP3 are speculated to interact with SARS CoV2 [14,15]. Additionally, the activation of the transcription factor NFκB and the NLRP3 inflammasome complex results in an enhanced expression of chemokines and pro-inflammatory cytokines such as (IL-2), IL-7, IL-6, granulocyte-colony stimulating factor (G-CSF), interferon- γ-inducible protein 10 (IP-10), monocyte chemoattractant protein-1 (MCP-1), macrophage inflammatory protein1 alpha (MIP-1A), interleukin-1 β (IL-1β) and Tumour Necrosis Factor α (TNFα) leading to a condition known as a cytokine storm [13] (Figure 3). This results in massive recruitment of immune cells such as neutrophils and monocytes to the alveolar lumen which further exacerbates the situation by producingmore ROS, reactive nitrogen species (RNS) and other toxic mediators like proteases and neutrophil extracellular traps (NETs) causing severe damage to the alveolar epithelium. The damaged epithelium leads to subsequent accumulation of oedematous fluid in the alveolar lumen resulting in hypoxemia and decreased carbon dioxide excretion which manifests as acute respiratory distress syndrome (ARDS) in the patients. While a majority of the people infected by SARS-CoV-2 virus recover quickly, few patients might develop ARDS and require hospitalization.

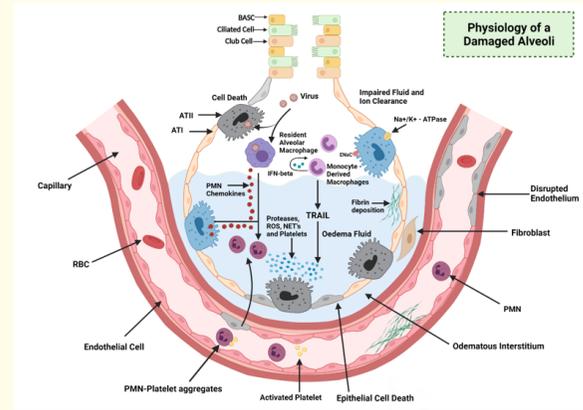

**Figure 2a:** Our alveolar epithelium is a continuous monolayer made up of alveolar type I (ATI) cells, which are very thin cells that allow for gas exchange, and alveolar type II (ATII) cells, which are primarily responsible for producing surfactant molecules that allow for lung expansion with low surface tension. By aiding the transfer of ions and fluid from the alveolus, both cell types help to maintain dry airspaces inside the alveolus. The intercellular tight junction connects the entire alveolar epithelium and is principally responsible for barrier function and regulating fluid and ion transport across the epithelium. Endothelial cells are the cells that make up the alveolar capillary and are in charge of regulating fluid input and the entry of inflammatory cells into the interstitial space. Intercellular junctions, which include both tight and adherens junctions, join these endothelial cells. Small solutes and water are moved into the interstitial space and then into the lymphatic system by the transvascular flux of fluid out of the alveolar capillary in normal circumstances. This fluid cannot cross the epithelial cell barrier in healthy normal alveoli, and resident alveolar macrophages inhabit the alveolar airspaces, providing host defence. Fibroblasts are present in the interstitial space, whereas PMNs and RBCs are situated within the alveolar capillaries. During the commencement of any infection, these PMNs can quickly migrate to the airspaces to carry out an immune response. [Created with BioRender.com.].

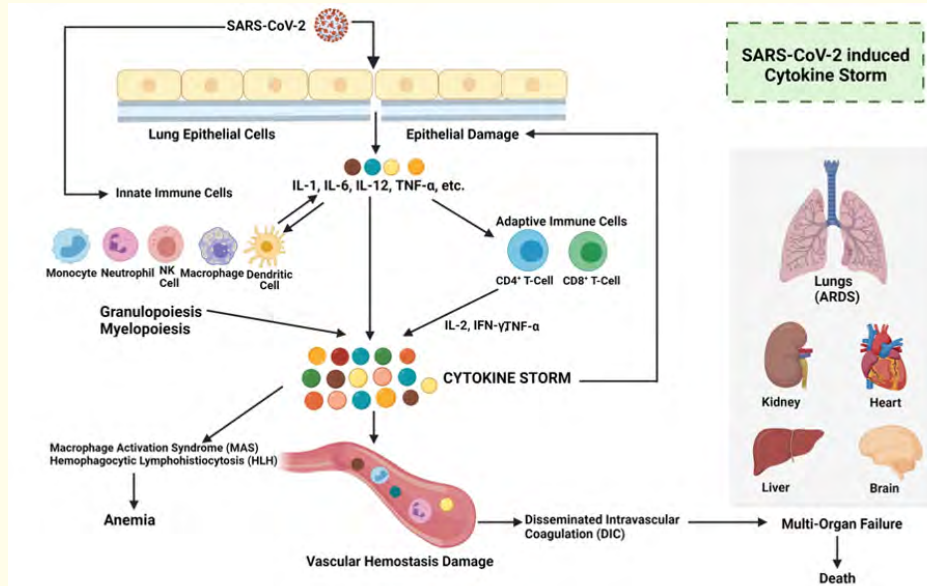

**Figure 3:** COVID- induced Cytokine Storm's immunopathological processes are depicted here. SARS-CoV-2 infects the host's epithelial cells or immune cells, causing tissue damage and the secretion of inflammatory cytokines (IL-1, IL-6, IL-12, and TNF) by epithelial cells and immune cells. These inflammatory cytokines activate adaptive immune cells (such as CD4+ T cells and CD8+ T cells) and recruit various innate immune cells (such as monocytes, macrophages, neutrophils, dendritic cells, and natural killer cells) to induce myelopoiesis [the production of bone marrow and the cells that arise from it] and emergency granulopoiesis [the production of granulocytes]. Apart from that, the activation of immune cells is linked to the creation of persistent and excessive circulating cytokines referred to as the Cytokine Storm, which might exacerbate epithelium injury. Furthermore, due to the triggering of Macrophage activation syndrome [MAS] and Hemophagocytic lymphohistiocytosis, the overproduction of systemic cytokines causes anaemia and disrupts vascular hemostasis (HLH). Most often, capillary leak syndrome, thrombosis, and DIC are the end results during such a condition. These occurrences combine to cause difficulties in the host's organs mainly the Lungs, Liver, Kidney, Heart, and Brain. All of these problems lead to ARDS, multiorgan failure, and death. [Created with BioRender.com.].

### ARDS in COVID 19 and pulmonary fibrosis

ARDS is thought to evolve in three coinciding phases: the exudative phase, the proliferative phase, and the fibrotic phase [16]. The exudative phase is defined by the release of pro-inflammatory cytokines. This is followed by an influx of immune cells into the alveolar lumen and destruction of the alveolar epithelial barrier resulting in respiratory distress due to alveolar edema. The fibroproliferative phase follows the exudative phase and is characterized by the accumulation of fibroblasts, fibrocytes, and myofibroblasts in the alveoli leading to an unwarranted deposition of fibronectin, collagen, and other matrix components [17]. This makes the lung extracellular matrix stiffer and hampers the lung microenvironment, eventually leading to dysregulation of the crosstalk across different components and cells. This is known to cause irreversible pulmonary fibrosis, characterized by scarring of the alveolar compartments and breathing difficulties.

However, in COVID-19 patients, the alveolar epithelium upon injury exhibits rapid proliferation and differentiation of replacement cells which can restore its barrier function. This has previously been studied for acute lung injuries due to bacterial infections [18]. Similar conclusions can be drawn for SARS-CoV-2 infection. In some cases, the stimulus is rapidly cleared and the basement membrane remains undamaged. Re-epithelization can occur to restore the lung to its previous state with minimal to no fibrosis. This restoration of the lung to its original histoarchitecture is termed as resolution [19] and might not require hospitalization (Figure 4).

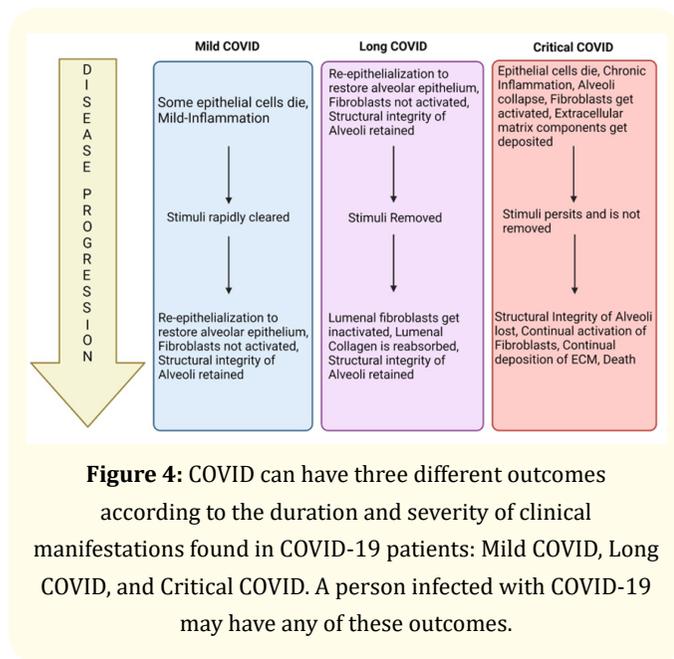

**Figure 4:** COVID can have three different outcomes according to the duration and severity of clinical manifestations found in COVID-19 patients: Mild COVID, Long COVID, and Critical COVID. A person infected with COVID-19 may have any of these outcomes.

In some cases, the host's immune system can clear out the SARS-CoV-2 virus at the initial stages of the fibroproliferative phase of ARDS, ensuring that the elastic structure and basement membrane of the alveoli are still intact. In such cases, re-epithelialization is possible concurrently with the breakdown of the collagenous tissue that has accumulated in the alveolar lumen by the fibrinolytic system. This scenario reflects 'resolving ARDS' [18] and is a plausible explanation for 'long-COVID'. These patients continue to experience discomfort and symptoms even post-recovery of a SARS-CoV-2 infection [19]. In cases of extensive damage to the lung alveoli, return to its previous histoarchitecture is improbable and precluded. This is marked by increasing amounts of fibrin and other matrix components' deposition alongside scar-tissue development. The resulting damage could either be diffuse or localized. This is possibly brought on by the disruption of host fibrinolytic system.

The primary component of the host fibrinolytic system is Plasmin, a serine protease. It is created when the precursor plasminogen is activated. Tissue-type plasminogen activator (tPA) and urokinase-type plasminogen activator (uPA) are the two activators in humans. The fibrinolytic system is controlled by serine protease inhibitors (serpins) at multiple activation sites to maintain a homeostatic equilibrium. The most effective inhibitors of plasmin are α-antiplasmin [20] and plasminogen activator inhibitor 1 (PAI-1) [16]. The tPA protein controls fibrin production in the blood. The uPA and its receptor uPAR play a major role in acute lung damage [17]. During transmission, the renin-aldosterone-angiotensin-system (RAAS) is disrupted by SARS-spike CoV-2's protein, which binds to the natural receptor angiotensin-converting enzyme 2 (ACE 2) in host cells. The RAAS is strongly associated with tPA and PAI-1.

Autopsy specimens of COVID-19 lungs exhibited diffuse alveolar destruction, capillary congestion, and capillary microthrombi [13]. Complement components [14] and neutrophil extracellular traps (NETs) were identified to be associated with pulmonary capillary microthrombi. D-dimer is formed when fibrin is broken down by proteolytic enzymes [15]. The D-dimer concentration has been linked to the severity of COVID-19 [15]. Angiotensin II upregulates PAI-1 in endothelial cells [16], which leads to reduced fibrinolysis and contributes to a hypercoagulable state. This indicates altered fibrinolysis in the pathogenesis of COVID-19. Scientists have disagreed about whether the COVID-19 patients' ARDS can be linked to the Berlin criteria of typical ARDS [20]. In COVID-19 patients, the exhibition of dry cough with very little sputum may suggest respiratory distress primarily due to damage to the alveolar epithelium. However, the endothelial cells remained less harmed with lesser exudation. Although ARDS- related respiratory failure is the most prevalent cause of hospitalization and mortality in COVID-19 patients [13], research on its molecular mechanism of disease establishment is still incomplete.

## Can MSC(s) be a potential cure for COVID-19?

## Role of lung endogenous progenitor cells in the lung tissue repair:

Most organs and tissues in our body utilize stem cells or their endogenous progenitor cells for repair. However, the lungs have an inherent regenerative capacity which increases when the lung encompasses severe damage. Various types of lung's endogenous epithelial progenitor and stem cells are involved in this process and help in the regeneration of damaged tissues [21]. The basal cells, which are the major progenitor cells in the epithelial lining of bronchioles migrate to the damaged alveolar epithelium and help in its repair [21,22]. Club cells, present in the distal part of the lung, also have self-renewal capacity. The terminal part of the tract consists of alveolar progenitor cells of ATII that eventually differentiate into ATII and ATI cells [22]. Single-cell analysis experiments have identified increased levels of Wnt-responsive ATII progenitor cells (TM4SF1+) and KRT5+ progenitor-derived alveolar barrier cells(ABCs) in COVID 19 patients. The former helps in the formation of ATII cells and ATI cells whereas the latter helps in repairingthe damaged alveolar epithelial lining (Figure 5).

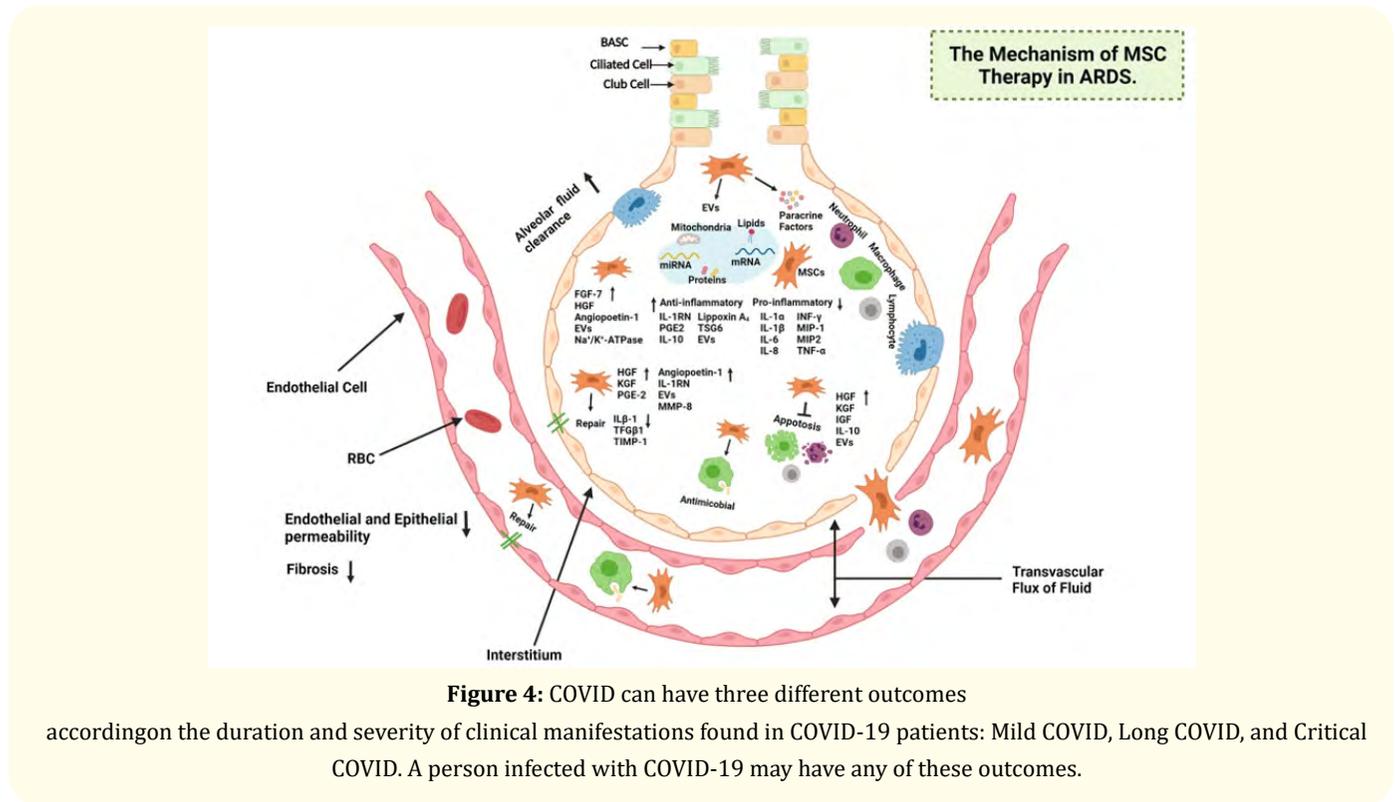

**Figure 4:** COVID can have three different outcomes accordingon the duration and severity of clinical manifestations found in COVID-19 patients: Mild COVID, Long COVID, and Critical COVID. A person infected with COVID-19 may have any of these outcomes.

### Therapeutic potential of mesenchymal stem cells (MSC(s)): What are MSCs?

Mesenchymal Stem cells(MSCs) are multipotent cells that have a mesodermal and ectodermal origin which has the capability to differentiate into several cell types of mesenchymal origin, including lung and alveolar epithelial cells, vascular endothelial cells and can play a major role in the regeneration of these cells if damaged by insults [23]. The major sources of MSCs are bone marrow, placental stem cells, adipose tissues, and the cells present in the blood of the umbilical cord [24]. MSCs were derived from bone marrow cells for the first time in 1978 by Friedenstein and team. They were called colony forming units (CFU) because of their ability to form *in vitro* colonies. They were soon named Mesenchymal stem cells on the basis of their mesenchymal origin during embryonic development in 1991 by Caplan., *et al.* [23]. In 2005, the International Society of Cellular Therapy (ISCT) termed these cells as multipotent mesenchymal stem cells and they laid down certain criteria for a cell to fall under this category.

### Immunomodulation by MSCs

*In vivo* studies have shown that MSC(s) possess the ability to suppress T cell activation and its proliferation by modulating several essential cytokines (Figure 5). MSC(s) can downregulate the secretion of IFN-γ and TNF- α and increase the amount of IL-4 in T cells. This triggers the transformation of a pro-inflammatory T cell into an anti-inflammatory state. Moreover, MSC(s) are implicated in the decrease of neutrophils in COVID 19 patients. They also increase the release of IL10 which can hamper the process of recruitment of neutrophils in the lungs [6,7].

The expression of HLA class II molecules and HLA class I molecules is absent in MSCs. They also lack the expression of various co-stimulatory molecules like CD80, CD86, CD40, and CD40L. This helps them overcome the cytotoxic effects of T and B cells present in the lymphatic system. Thus they are also termed as 'immune-privileged cells' [25].

### Tissue repairing and Anti-apoptotic capability of MSC(s)

MSC(s) possess self-renewal capacity as well as the potential to differentiate into different tissue types. Apart from this, the therapeutic effect of MSC(s) increases because of the secretion of important cytokines and growth factors that are responsible for extensive tissue regeneration and repair [26]. Some of the important factors secreted by them are the keratinocyte growth factor (KGF), vascular endothelial growth factor (VEGF), and hepatocyte growth factor (HGF). They can reduce the amounts of TGF-**β** and TNF-**α** which also help in reducing the collagen content. MSC(s) are also known to secrete certain alveolar surface- active substances which in turn increases the regeneration of the epithelial cells and increases angiogenesis [8]. The recruitment of immune cells in the lungs during an insult is linked with apoptosis of the lung cells. In SARS CoV2 patients, especially those who are severely affected are known to have a condition of lymphocytopenia (excessive apoptosis of the lymphocytes) which in turn impairs the CD4+T cells [27]. There are also reports of severe alveolar epithelial and endothelial cell apoptosis [28]. MSC(s) induced secretion of keratinocyte growth factor (KGF), angiopoietin-1, and hepatocyte growth factor (HGF) has the capability to reduce alveolar epithelial and endothelial apoptosis [29]. MSC(s) are also known to reduce apoptosis in macrophages induced by Lipopolysaccharide (LPS) by inhibiting the Wnt/β-catenin pathway [30].

### Antiviral and antibacterial properties of MSC(s)

MSC(s) are known to reduce viral replication and virus shedding in epithelial cells of the lungs and they have shown extensive potential in reducing the severity of diseases caused by various viruses, including Herpes Simplex Viruses, Influenza viruses, cytomegaloviruses, measles viruses, and hepatitis C viruses. IFN-Y induced production of indoleamine 2,3-dioxygenase (IDO) and secretion of LL37 by MSC(s) has the capability of some of the viral membrane degradation as well as attenuation of replication of several viruses [31]. Reports have been there about the anti-inflammatory and anti-viral properties of MSC-derived extracellular vesicles which have the capability of transferring RNA(s) to lung epithelial cells. MSC(s) can also secrete many antibacterial and antimicrobial peptides and proteins, including LL-37, lipocalin-2, and βdefensin-2, hepcidin, and KGF that helps in bacterial clearance [32,33]. The MSC(s) are also reported to increase the phagocytic activity of macrophages and monocytes by mitochondrial translocation via nanotubes in ARDS models [32].

### Ability to protect pulmonary endothelial and alveolar epithelial cells as well as increasing alveolar fluid clearance

Excess alveolar and interstitial fluid in the lungs hinders surfactant concentration and gas exchange, there fore eliminating it promotes functional recovery in lungs. This is achieved by secreting HGF through extracellular vesicles, lowering inflammatory damage, and promoting autophagy. MSC treatment has been further proven to protect or repair the alveolar epithelial and pulmonary endothelial lining. MSc therapy has also been proven to protect alveolar epithelial cells from inflammation and oxidative stress by secreting angiopoietin-1, interleukin 1 (IL-1) receptor antagonist (IL-1RN), prostaglandin E2 (PGE2), HGF, and KGF, or scavenging oxidants and radicals [34]. Furthermore, in an animal model of ARDS, MSCs can control tissue regeneration and repair processes, reduce lung fibrosis by boosting metalloproteinase (MMP)-8, reduce tissue inhibitors of metalloproteinase (TIMP)-1, IL-1, and transforming growth factor-1 (TGF-1) levels.

MSc therapy eliminates interstitial fluid in a lung damage model produced by influenza infection by secreting angiopoietin-1 and KGF and inhibiting the downregulation of Na+/K+ -ATPase [35]. Screening of bronchoalveolar lavage fluid from COVID-19 patients, have shown increased viral load of SARS-CoV-2 virus [2], and highly inflammatory monocyte-derived FCN1+ macrophages are present and are responsible for production of large amount of cytokines, indicating an inflammatory niche in the alveolar fluid. As a result, MSC transplantation may help patients with COVID-19 regain lung function by increasing alveolar fluid evacuation.

### Genetic modification and preconditioning of MSCs

MSC therapy comes with its own challenges. It has been observed that MSC(s) thrive to survive after administration which is also a result of their lower migration to bone marrow. This affects their homing and proliferation. Genetic modification or changing

the genetic makeup of the MSC(s) to make them more effective and as a result enhance their survivability. Strategies have been used to overexpress the beneficial genes in MSC therapy previously as a potential therapeutic model for ARDS. These over-expressed factors have shown improved results in MSC therapy of ARDS and in some LPS-induced ALI models of lungs (Table 1). However, genetic modification of MSCs may lead to genome instability and other associated risks. Thus, alternatives have been designed. Preconditioning strategies to improve MSC(s) viability have shown promising results in ARDS (Table 2). Modifying MSC(s) with these factors or using preconditioning strategies will play important roles in employing MSC therapy for SARS-CoV-2 treatment. Nonetheless, extensive research is required to confirm their efficiency.

| Genetic modification(s) | |
|---|---|
| **Factors:** | **Functions:** |
| E-prostanoid 2(EP2) receptor | Increases the migration of MSC(s) to damaged tissues when activated by PGE2released in damaged inflammatory sites [47]. |
| Angiotensin- converting enzyme2 (ACE2) | Increases migration of MSC(s), degrades angiotensin II which plays a role in fibrosisand is a profibrotic peptide and restores endothelial function in lung injury [48]. |
| Hepatocyte growthfactor (HGF) | Restores the human lung endothelial cells permeability, protects the lung from insults, Increases MSC(s) viability and survival in the lung tissue, decreases the pro-inflammatory phenotype of MSC(s) in stressed conditions [48]. |
| Keratinocyte Growth Factor(KGF) | Repairs pulmonary epithelial cells, Downregulates pro-inflammatory responses and enhances anti-inflammatory responses in the lung, increases microvascular permeability, Increases type II epithelial cells proliferation, and increases surfactants inlungs [49]. |
| Heme oxygenase-1(HO-1) | Possesses anti-inflammatory, anti-apoptotic, anti-oxidative properties. Helps MSC(s) toneutralize the cytokine storm and to maintain homeostasis. Checks and modulates neutrophil count and protein concentration in bronchoalveolar fluid ameliorates the structure of tissue architecture in damaged lungs [50]. |

**Table 1:** Genetic modifications in MSC(s) for increased efficiency.

| Genetic preconditioning | |
|---|---|
| **Strategies** | **Functions** |
| Hypoxic pre-treatment or conditioning | Increases the secretion of pro survival and proangiogenic factors, reduces apoptosis, induces the expression of anti-apoptotic, anti-oxidative, and proangiogenic factors like vascular endothelial growth factor (VEGF), angiopoietin 1, HGF, insulin-like growthfactor 1 (IGF-1), and bal-2 [51]. |
| Serum obtained from ARDS patients | Enhanced anti-inflammatory properties of MSC(s) preconditioned with ARDS serumby secretion of IL10 and IL-1-RN, reduction of pulmonary edema and lung injury [52]. |
| TGF-β | Enhance the concentration of ECM components like fibronectin, increases the survivalof MSC(s), role in tissue repair, and prolonged effect of MSC therapy [53]. |

**Table 2:** Different preconditioning strategies to increase viability of MSC(s).

### Why choose MSCs as a cure for COVID 19 ARDS?

Unchecked stimulation of the host's innate immune system during a SARS-CoV-2 infection can often lead to an excessive build-up of pro-inflammatory cytokines resulting in a cytokine storm. This can result in acute respiratory distress syndrome (ARDS) and a myriad of other disastrous consequences which may include paving the pathway for secondary infections in the host [36]. A point of great importance that has often been neglected is the efforts to mitigate the effects of the disease on recovered patients is that most of them may have the effects of COVID-19 for years and even the rest of their lives in addition to suffering irreversible damage [37]. There are sufficient instances of irreversible pulmonary, cardiac, and renal impairment in recovering patients. Subsequently, some of the drugs presently used in the treatment of SARS-CoV-2 include anti-inflammatory remedies such as NSAIDS, TNF inhibitors, JAK inhibitors, glucocorticoids, etc. [38] which are aimed at modulating the cytokine storm. However, the difficulties posed in the treatment of severe ARDS continue to remain. As such, scientists all over the world are looking for alternate treatments for countering the cytokine storm in a more effective manner. Studies, including several clinical trials, have indicated that stem-cell-based therapies may be the way forward [36].

### Clinical trials of MSc therapy in COVID 19 performed to date

Current treatment options for COVID-19 patients mostly include re-purposed antiviral and immunomodulatory drugs. However, accumulating evidence suggests that severely or critically ill COVID-19 patients would require adjuvant therapies in addition to the current treatment options for holding a better chance at survival and recovery [39]. Presently, the adjuvant therapy options mostly employed include regenerative-medicine-based therapies which involve infusing the patients with convalescent plasma. While effective, the problem with convalescent plasma therapy lies in its severe dependence on plasma collection programs at local demographic levels from fully-recovered COVID-19 patients. Alternatively, regenerative-medicine-based therapies dependent on MSCs or MSC-derived extracellular vesicles (EVs) appear to be much more lucrative in terms of scalability and distribution capacity [39].

In order to locate mesenchymal stem cell therapies undergoing clinical trials for use against COVID-19, the ClinicalTrials.gov database (https://clinicaltrials.gov/ct2/home) was searched. In addition, we also looked up the clinical trials on mesenchymal stem cell therapies against COVID-19 in the World Health Organization's International Clinical Trials Registry Platform (WHO ICTRP). The WHO ICTRP platform consists of clinical trials registered in databases maintained by countries other than the United States of America. The figure below (Figure 6a, Figure 6b) shows the global distribution of the clinical trials being carried out to check the safety and efficacy of mesenchymal stem cells and their derivatives against COVID-19 (data as of June 2022).

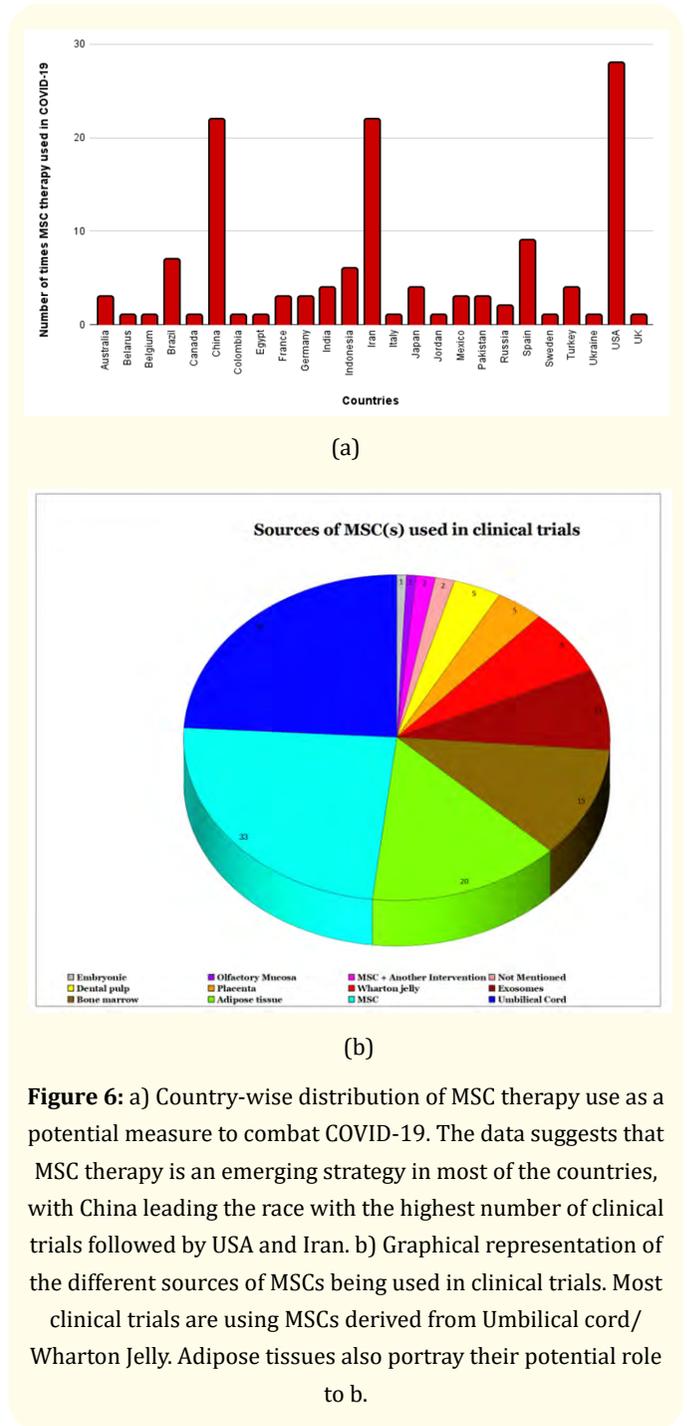

(a)

(b)

**Figure 6:** a) Country-wise distribution of MSC therapy use as a potential measure to combat COVID-19. The data suggests that MSC therapy is an emerging strategy in most of the countries, with China leading the race with the highest number of clinical trials followed by USA and Iran. b) Graphical representation of the different sources of MSCs being used in clinical trials. Most clinical trials are using MSCs derived from Umbilical cord/ Wharton Jelly. Adipose tissues also portray their potential role to b.

Mesenchymal stem cells (MSCs) can be obtained from numerous sources. For their ease of scaling up to clinical concentrations and their readily available cellular sources, a significant number of the clinical trials registered in public databases are employing/have employed mesenchymal stem cells derived from umbilical cords to use against COVID-19. Some of the earliest trials for testing the safety and efficacy of mesenchymal stem cells against COVID-19 have employed umbilical cord-derived-MSCs (UC-MSCs). In a study conducted by Feng., *et al*. [40] on 16 COVID-19 patients from February to the first of April, in early 2020, UC-MSCs were used. The patients enrolled in the study included 7 who had critically severe and 9 who had severe symptoms. The study reported promising results with improvement in the patients' oxygenation index, procalcitonin (PCT), CRP, etc.

### Problems and future directions

Despite the promising results being reported by mesenchymal stem cell-based clinical trials, the healthcare community continues to search for the exact mechanism of action of mesenchymal stem cells in COVID-19 patients. This, along with the unavailability of pre-clinical models of COVID-19 further hinders research into this area.

A major roadblock to MSC clinical trials is the low enrolment of COVID-19 patients. The spectrum of clinical manifestations of COVID-19 in patients and the range of immune responses make it difficult to predict the disease progression. Additionally, the long-term effects of this treatment are yet to be documented. However, this can be partially addressed by newer, more scientific methods of consultation such as Telemedicine. A study conducted by [41] suggests that telemedicine could be an equally effective way of monitoring patient progress. This new cost effective and time saving method could potentially help us evaluate the long-term effects of MSC therapy in COVID-19 patients.

Another practical limitation to MSC therapy is that there is no standardized protocol for it in COVID patients. This combined with the high cost and the necessity of highly trained staff and equipment for the administration makes it an expensive alternative. One other challenge associated with studying the effects of these therapies and visualizing the long-term effects in an *in vivo* condition is the lack of appropriate experimental setups which can mimic the human lung *in-vivo* environments. Recently, lung organoids and lung-on-a-chip strategies have proven their potential to recapitulate the variability, severity, and uncertainty associated with SARS-CoV-2 infection in human lung epithelial cells [42]. Lung organoid models have been derived from human embryonic stem cells (hESCs) and induced pluripotent stem cells (hiPSCs) which have increased the self-renewal and self-organizing capacities of the organoids *in vitro* [43]. The use of precise growth factors to induce iPSC differentiation into a specific cell lineage to simulate lung patterning has resulted in a better understanding of the COVID-19 lung disease [42,43].

Bio-inspired design of lung on-chip models has utilized multicompartment microfluidic channels in an attempt to imitate the human alveolar-capillary systems. The use of human pluripotent stem cells in this technique has been proposed to make this strategy cost-effective. This also allows necessary gene editing or mutations to be performed, reducing the need for intrusive procedures such as a biopsy to study disease- related mutations [42]. Moreover, MSCs have enhanced distal lung epithelial cell differentiation, increasing the development of both alveolar and small airway epithelial cells in 3D lung organoids.

Although these techniques have shown potential, they also come with certain limitations. Producing uniformly sized lung organoids and maintaining physiological cell ratio during their self-organization is a challenging task. A lung on a chip system can address this constraint by cultivating cells in a uniform orientation, resulting in a more reproducible cell culture method. However, the use of PDMS membranes coated with ECM elements in a lung on a chip system, essentially the one which mimics the interface of tissues, has shown to have different transport and structural permeabilities or properties than the physiological condition or *in vivo* [42]. Overcoming these constraints and progressing these techniques in the near future will give us a useful system for visualizing the physiological, spatiotemporal, and long-term effects of the different COVID-19 treatments.

### Conclusion

In a recent study conducted by Hasan., *et al*. 2020, the overall mortality rate estimated among 10,815 ARDS cases in COVID-19 patients was almost 39% [44]. During this unprecedented healthcare crisis, researchers and doctors from all around the world are working tirelessly to discover the best methods and treatments for COVID-19 patients. The high fatality rates make immediate actions imperative, and MSCs might be the key to pre-empting these problems.

In this paper, we have highlighted the key differences between normal ARDS and COVID induced ARDS and found that MSCs might help reduce the long-term effects of the latter. Despite all of the difficulties connected with the treatment, MSC therapy has demonstrated positive outcomes in both lab and clinical studies when all other therapies have failed.

Mesenchymal stem cells have shown positive effects in patients with Influenza induced damage in lungs. Similarly, we posit that the use of MSC on COVID-19 ARDS patients has the potential to lower mortality rates. We can conclude that MSCs can be used as a plausible cure for long COVID and can be used to combat the lethal effects of COVID induced ARDS. MSC therapy could perhaps also be useful for the treatment of other COVID-19 induced complications such as subarachnoid haemorrhage (SAH), myocarditis, acute kidney injury, etc. which are brought on by the host's hyper active immune system [45,46]. However, further studies need to be conducted in order to determine the efficacy of MSC therapy on such COVID-19 induced complications. Standardizing MSC procedures and clinical application techniques is critical and a standardised protocol will make the treatment more available to critical patients. Further research needs to be done to understand the long-term effects of MSC therapy. MSc handling also necessitates the use of skilled individuals and complex equipment, but measures should be done to lower the cost of good treatment and make it more accessible to the general population.

## Funding

Not Applicable.

## Conflicts of Interest

The authors have no conflicts of interest to declare that are relevant to the content of this article.

## Ethics Approval

Not Applicable.

## Consent to Participate

Not Applicable.

## Consent for Publication

Not Applicable.

## Availability of Data and Material

Not Applicable.

## Code Availability

Not Applicable.

## Authors' Contributions

All authors have contributed equally to this paper.